\documentclass[showpacs,preprintnumbers,amsmath,amssymb,twocolumn,superscriptaddress,prb]{revtex4}
\usepackage{times}
\usepackage{graphicx}
\usepackage{amsfonts}
\usepackage{amsmath, amsthm, amssymb}
\usepackage{dsfont}
\usepackage{subfigure}

\newcommand{\vk}{\mathbf{k}}

\newcommand{\e}[1]{\mathrm{e}^{#1}}

\newcommand{\ie}{\textit{i.e. }}%[syn: f.eks., for example, for instance]
\newcommand{\eg}{\textit{e.g. }}%[syn: f.eks., for example, for instance]
\newcommand{\etal}{\emph{et al. }}
\def\i{\mathrm{i}}

\begin{document}
\title[Josephson current in diffusive multilayer superconductor/ferromagnet/superconductor junctions]{Josephson current in diffusive multilayer superconductor/ferromagnet/superconductor junctions}
\author{Iver B. Sperstad}
\affiliation{Department of Physics, Norwegian University of
Science and Technology, N-7491 Trondheim, Norway}
\author{Jacob Linder}
\affiliation{Department of Physics, Norwegian University of
Science and Technology, N-7491 Trondheim, Norway}
\author{Asle Sudb{\o}}
\affiliation{Department of Physics, Norwegian University of
Science and Technology, N-7491 Trondheim, Norway}

\date{Received \today}
\begin{abstract}
We calculate the Josephson current in a diffusive superconductor/ferromagnet/superconductor junction, where the ferromagnetic region 
contains multiple layers (or domains). In particular, we study a configuration where there are two layers with an arbitrary relative 
in-plane magnetization orientation, and also include non-ideal interfaces and arbitrary spin-flip scattering. We study the 0-$\pi$ 
oscillations of the critical current for varying junction width $d$, and find that the $\pi$ state vanishes entirely when the 
magnetic misorientation angle of the two layers exceeds a critical angle $\phi_c$. While $\phi_c \to \pi/2$ in the limit of high 
temperatures, we find that $\phi_c$ becomes smaller than $\pi/2$ at low temperatures compared to $T_c$. 0-$\pi$ oscillations are also found when
varying the temperature or the misorientation angle for fixed values of $d$, and we present phase diagrams that show
qualitatively the conditions for the appearance of such oscillations. We also point out how one may obtain significant
enhancement of the critical current in such a system by switching the magnetization for selected values of the junction
width $d$, and comment on the necessary conditions for establishing a long range triplet Josephson effect. 

\end{abstract}

\maketitle

\section{Introduction}

The topic of superconductor-ferromagnet heterostructures has been a subject of intense research for several years \cite{bergeretRMP, buzdin}. 
Not only do such systems constitute model systems for investigating the interplay between two fundamental condensed matter 
phenomena, ferromagnetism (F) and superconductivity (S), but recent advances in fabrication techniques of such hybrid 
structures make applications increasingly attainable. Especially S/F based Josephson technology holds great promise in 
nanoelectronics, \eg as a physical realization of the qubit of quantum computation\cite{qubit}. Another possible device 
is in some sense analogeous to a spin valve exhibiting giant magnetoresistance (GMR), \ie strongly suppressing the current
for opposite orientation of the magnetization \cite{gmr} in two F layers separated by a normal metal(N). For our object 
of interest however, superconducting electrodes are used instead of ferromagnets and N is replaced with F, in which 
case the resistance effect of magnetization-switching is known to be reversed compared to the spin valve\cite{bergeretJOS}.

The proximity effect between a superconductor and a normal metal was predicted decades ago \cite{proximity} and has since 
been investigated thoroughly both theoretically and experimentally. However, several new and interesting effects were 
predicted when the layer of normal metal was replaced with a ferromagnet, due to spin-triplet correlations in the 
ferromagnet induced by the exchange field \cite{bergeretRMP,buzdin}. Much attention has been given to SFS structures, 
which are studied as a somewhat more exotic class of Josephson junctions. The most interesting emerging phenomenon in 
SFS junctions is the appearance of the so-called $\pi$-state \cite{bulaevskii, buzdin_bulaevskii}, in which the 
difference in the superconducting phase across the junction is $\pi$ in the ground state, in contrast to the conventional 
state with phase difference zero. The physical result of transitions between these states is usually a sign change in 
the critical Josephson current $I_\mathrm{c}$ through the junction, the observable manifestation of which being a 
non-monotonic dependence of $I_\mathrm{c}$ on parameters such as temperature and junction width. Been predicted for 
decades, the experimental verification of this phenomenon some years ago \cite{ryazanovPRL01, kontos} was one of the 
catalysts of the present activity on the field. Recently, the effect of magnetic impurities in SFS junctions was also investigated theoretically \cite{gusakova, bergeret3}.

It is well known that in simple S/F structures, the proximity effect will only induce OSP triplet correlations (Opposite 
Spin Pairing, spin projection $S_z = 0$), and that ESP triplet correlations (Equal Spin Pairing, $S_z = \pm 1$) require inhomogeneous magnetization \cite{bergeretRMP}. ESP components are in some contexts referred to as LRTC components 
(Long Range Triplet Correlations), and are of special interest because they do not decay as rapidly in the ferromagnet 
as the other components, and may therefore evade the suppression of the supercurrent for increasing width of the 
ferromagnet. One way of achieving this in theory is to let the magnetization in the F layer have a helical 
structure \cite{bergeret_spiral, eschrig1}. A similar effect is also found by considering a SIS junction where 
the S electrodes themselves exhibit a spiral magnetic order coexisting with the superconductivity \cite{kulic}. 
Yet another alternative is layered S/F structures \cite{bergeret_layers} with noncollinear magnetization, or 
simply a SFS structure where F is replaced with several ferromagnets with different direction of magnetization.

Recent theoretical studies have been focusing on junctions with two ferromagnets sandwiched between the superconducting electrodes. 
There are two physical realizations of a system described by such a model. It may either describe a device with two distinct, 
consecutively placed F layers (a SF$_1$F$_2$S junction), constructed to achieve customized nonhomogeneous magnetization. Several 
works \cite{pajovic, barash, bergeretJOS, linderjos,belzigjos,li2002, eschrig2} have considered the 
Josephson current in such heterostructures. On the other hand, the model may describe the more realistic experimental situation 
of one ferromagnetic layer with several magnetic domains. Some models of the latter kind have included two or more in-plane 
magnetic domains \cite{fominov_neel,crouzy2}, and although this certainly is an interesting framework for understanding real 
heterostructures, the present paper considers the two layers placed consecutively as a SF$_1$F$_2$S structure. The main 
motivation for this choice is that this configuration allows for much easier experimental control of the magnetization. 
The misorientation angle may be tuned by applying an external weak magnetic field to the interface between the ferromagnets 
if the magnetization axis is pinned in one layer, while in the other one there is an in-plane easy axis. \cite{barash} To accomplish this experimentally, one would probably need some interlayer between the ferromagnetic films to avoid a locking between the corresponding magnetizations due to the interfilm exchange coupling. 

Calculations on the models referred to here have predicted 0-$\pi$ transitions upon varying the strength of the magnetic 
exchange field, the junction width or the temperature, depending on the relative orientation of the magnetization in the 
two F layers \cite{pajovic, barash, bergeretJOS, linderjos,belzigjos,li2002, eschrig2}. For 
antiparallel magnetization, it is reported that the 0-$\pi$ oscillations will vanish, rendering the $\pi$ state impossible 
\cite{blanter_hekking}.  An enhancement of the critical current for the antiparallel orientation by increasing the 
exchange field was first reported by Bergeret \etal \cite{bergeretJOS}, and shortly after elaborated upon by others
\cite{golubovJETP,koshina}. Much of the work has however been limited to the case of collinear magnetization 
\cite{blanter_hekking, koshina}, but recently also SF$_1$F$_2$S systems with arbitrary misorientation angle for 
the magnetization have been analysed \cite{crouzy1, pajovic, barash}. In particular, Crouzy \etal have shown 
\cite{crouzy1} how the $\pi$ state of such a junction is suppressed for increasing misorientation angles, 
vanishing at a critical angle $\phi_c = \pi/2$ when the temperature is close to the critical temperature, 
\ie $T/T_c \simeq 1$. One recent article \cite{trilayer} has even studied a corresponding ferromagnetic 
trilayer structure, but focused chiefly on the LRTC contribution to the Josephson current in such a 
structure. It should be noted that while the majority of the work in this field is carried out in the 
dirty limit, considering diffusive F/S systems, several of the relevant papers 
\cite{barash, radovic, pajovic, trilayer} study ballistic junctions as well.

Ref.~\onlinecite{crouzy1} points out the necessity of including additional effects to get a more accurate description of such 
systems. The present article may thus be viewed as an extension of their work by studying a SF$_1$F$_2$S junction with 
noncollinear domains where non-ideal interfaces and magnetic impurities are also taken into consideration. For the latter, 
we will study the special cases where isotropic or uniaxial spin-flip scattering is present. Consequently, there are three 
questions addressed in this work which were not treated in Ref.~\onlinecite{crouzy1}: \textit{i)} how does spin-flip scattering influence the 0-$\pi$ oscillations, \textit{ii)} do nonideal 
interfaces change the qualitative behaviour of the system, and \textit{iii)} how does the Josephson current for such a system 
depend on the temperature. The possibility of investigating the latter point is present in our model, as opposed to 
Ref.~\onlinecite{crouzy1}, which was restricted to temperatures close to $T_c$. The reason for this is that the regime 
of weak proximity effect is only attainable either if the S/F transparency is low, or when transparency is high under 
the extra restriction that $T \approx T_c$, so that it is guaranteed that the influence of superconductivity is weak 
in either case.

\begin{figure}[h!]
\centering
\resizebox{0.46\textwidth}{!}{
\includegraphics{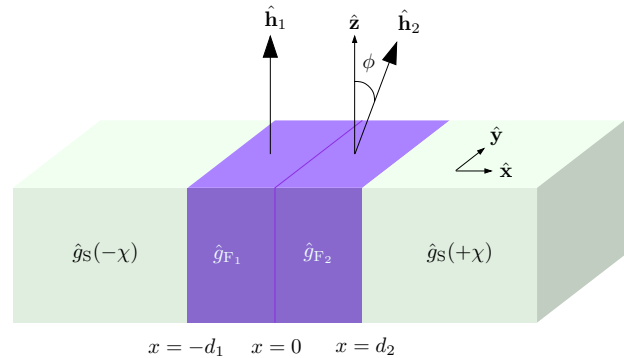}}
\caption{(Color online) The experimental setup proposed in this paper. Two $s$-wave superconductors are separated by two 
ferromagnetic layers with an arbitrary relative orientation of the magnetizations. The ferromagnetic regions may model a 
domain-structure of a single ferromagnetic layer or correspond directly to two distinctly deposited magnetic layers. This 
is similar to the setup considered in Ref.~\onlinecite{crouzy1}.}
\label{fig:model}
\end{figure}

For concreteness, we consider a diffusive Josephson junction with two ferromagnetic layers with arbitrary in-plane relative 
orientation of the magnetization, as shown in Fig. \ref{fig:model}. Although we focus on this picture of distinct, controllable 
layers, the physically similar situation of magnetic domains will also be commented upon. The superconducting electrodes are 
two similar s-wave superconductors, and the interfaces between the superconductors and the ferromagnet are assumed to have 
low transparency.

This paper is organized as follows. In Sec. \ref{sec:theory} we will briefly sketch the theoretical framework and go on to obtain 
a solution for the proximity-induced anomalous Green's function in the ferromagnetic region of our SF$_1$F$_2$S system, from which 
an expression for the Josephson current can be calculated. In Sec. \ref{sec:results} we present the dependence of the Josephson 
current on the various parameters, analyse the resulting 0-$\pi$ oscillations in the junction and construct a corresponding phase 
diagram. We discuss the applicability of our findings in Sec. \ref{sec:disc}, and furthermore present a discussion of the absence 
of the long range Josephson effect in such SF$_1$F$_2$S systems. A summary and some final remarks are given in 
Sec. \ref{sec:summary}. We will use boldface notation for 3-vectors, $\hat{\ldots}$ for $4\times4$ matrices, and 
$\underline{\ldots}$ for $2\times2$ matrices.

\section{Theoretical formulation}\label{sec:theory}
We address this problem by means of the Usadel equation in the quasiclassical approximation. This corresponds to integrating 
out the dependence on the kinetic energy of the Gor'kov Green's function, obtaining thus the quasiclassical Green's function 
$\hat{g}$ as the object used to describe our system. This approximation is valid as long as all relevant energy scales are much 
smaller than the Fermi energy $\varepsilon_\mathrm{F}$, and correspondingly that all relevant length scales are much larger 
than the Fermi wavelength. The latter condition is reconciled with the presence of sharp interfaces in our model by introducing 
appropriate boundary conditions, as discussed below. The approach is based on the Keldysh formalism for nonequilibrium 
superconductors, which is convenient to work in also in the present limiting case of equilibrium. Here one operates with 
$8\times8$ matrix Green's functions, in which the retarded Green's function $\hat{g}^R$, the advanced Green's function 
$\hat{g}^A$ and the Keldysh Green's function $\hat{g}^K$ are $4\times4$ matrix components. As both the advanced and the 
Keldysh component of the matrix can easily be expressed by the retarded component in our case, it will be implicitly 
assumed in the following that the Green's function under consideration is the retarded component $\hat{g}^R$.

We will desist from further discussion of quasiclassical theory, and instead refer the reader to the considerable literature 
that covers the Keldysh formalism and nonequilibrium Green's functions \cite{serene, rammer, zagoskin, belzigreview,jpdiplom}. 
We go on to write up the matrix structure of our quasiclassical Green's functions. In the bulk superconductors, the Green's 
function reads \cite{serene}
\begin{equation}
\hat{g}_S = \begin{pmatrix}
\underline{1}c & \i\underline{\tau_2} s\e{\pm\i\chi} \\
\i\underline{\tau_2}s\e{\mp\i\chi} & -\underline{1}c\\
\end{pmatrix},
\end{equation}
where $c \equiv \text{cosh}(\theta)$, $s \equiv \text{sinh}(\theta)$, and $\theta \equiv \text{atanh}(\Delta_0/\varepsilon)$, 
with $\Delta_0$ denoting the amplitude of the superconducting gap. The different signs of the phase $\chi$ above correspond 
to the left (lower sign) and right (upper sign) superconducting bank, respectively. When not being in proximity to a 
superconductor, the Green's function for a bulk ferromagnet reads
\begin{equation}
	\hat{g}_{F,0} = \begin{pmatrix}
	\underline{1} & 0 \\ 
	0 & -\underline{1}\\
\end{pmatrix}.
\end{equation}
When being influenced by a superconductor, off-diagonal elements is introduced to this Green's function, and for weak 
proximity effect it is changed to $\hat{g}_F \approx \hat{g}_{F,0} + \hat{f}$. This perturbation can be expressed 
as \cite{serene}
\begin{equation}
	\hat{f} = \begin{pmatrix}
		0 & \underline{f}(\varepsilon)\\
		-[\underline{f}(-\varepsilon)]^\ast & 0\\
	\end{pmatrix},
\end{equation}
where the constituting anomalous Green's function can be written as a matrix in spin space on the form \cite{leggett1975}
\begin{equation}
	\label{eq:anomaleous}
	\underline{f} = \begin{pmatrix}
		f_\uparrow & f_\text{t}+f_\text{s}\\
		f_\text{t}-f_\text{s} & f_\downarrow\\
	\end{pmatrix}.
\end{equation}
Here, $f_\text{s}$ denotes the singlet component, $f_\text{t}$ the OSP triplet component and $f_\uparrow$ and 
$f_\downarrow$ the ESP triplet components, and it is these anomalous Green's function that the Usadel equation 
is to be solved for.

In our calculations, we will account for the possibility of both uniaxial and isotropic spin-flip scattering 
by the parameter $\eta_{xy}$ and $\eta_z$ as follows:
\begin{align}
&\text{Uniaxial spin-flip:}\; \eta_{xy}=0,\; \eta_z=3,\notag\\
&\text{Isotropic spin-flip:}\; \eta_{xy}=1,\; \eta_z=1.
\end{align}
The spin relaxation time for spin-flip scattering will be denoted $\tau_\text{sf}$, and is to be considered as 
a phenomenological parameter in our approach.

\begin{figure}[h!]
\centering
\resizebox{0.15\textwidth}{!}{
\includegraphics{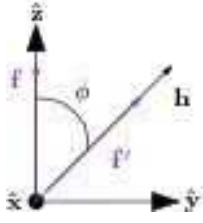}}
\caption{The change of spin-basis from a quantization axis $\parallel \hat{\mathbf{z}}$ to a quantization axis $\parallel \mathbf{h}$.}
\label{fig:basis}
\end{figure}

In the ferromagnetic regions F$_1$ and F$_2$, the linearized Usadel \cite{usadel} equations take the form
\begin{align}\label{eq:linusa1}
&D\partial_x^2 (f_\text{t}\pm f_\text{s}) + 2\i(\varepsilon\pm h\cos\phi)(f_\text{t}\pm f_\text{s}) \notag\\
&- \frac{1}{2\tau_\text{sf}}(\eta_zf_\text{t} \pm 3f_\text{s}) \pm h\sin\phi(f_\uparrow+f_\downarrow) = 0,\notag\\
&D\partial_x^2 f_\sigma + (2\i\varepsilon -\frac{\eta_{xy}}{2\tau_\text{sf}})f_\sigma - 2h\sin\phi f_\text{s} = 0;\; \sigma=\uparrow,\downarrow,
\end{align}
with $\phi=0$ in $F_1$.
Eqs. (\ref{eq:linusa1}) constitute a set of coupled, second-order, linear differential equations. Although an 
explicit analytical solution may be obtained for $\{f_{\text{s}},f_{\text{t}},f_{\sigma}\}$ by solving 
Eq. (\ref{eq:linusa1}) brute-force for non-zero $\phi$, the resulting expressions are very large and cumbersome. 
We therefore proceed via an alternative but equivalent route. By a change of spin basis to a quantization axis 
which is aligned to the exchange field in F$_2$, one obtains the equations
\begin{align}
&D\partial_x^2(f'_\text{t} \pm f_\text{s}) + 2\i(\varepsilon\pm h) (f'_\text{t} \pm f_\text{s}) - \frac{1}{2\tau_\text{sf}}(\eta_zf'_\text{t} \pm 3f_\text{s}) = 0,\notag\\
&D\partial_x^2 f'_\sigma + (2\i \varepsilon - \frac{\eta_{xy}}{2\tau_\text{sf}}) f'_\sigma = 0.
\end{align}
The superscript $'$ denotes the new spin basis, and the $s$-wave component transforms as a scalar under spin 
rotations: $f_\text{s}' = f_\text{s}$. The general analytical solution for these equations in the case of 
isotropic spin-flip scattering was obtained in Ref.~\onlinecite{linderUsadel07}. In the present case, we obtain
\begin{align}\label{eq:general}
f'_\text{t} &= c_1 \e{-q_-x} + c_2\e{q_-x} + c_3\e{q_+x} + c_4\e{-q_+x},\notag\\
f_\text{s} &= \frac{\i}{8\tau_\text{sf}h}(c_1\kappa_+ \e{-q_-x} + c_2\kappa_+\e{q_-x} \notag\\
&\hspace{0.5in}+ c_3\kappa_-\e{q_+x} + c_4\kappa_-\e{-q_+x}),\notag\\
f'_\sigma &= A_\sigma \e{\i kx} + B_\sigma\e{-\i kx},
\end{align}
where we have defined 
\begin{align}
	\label{eq:q_k}
q_\pm &= \sqrt{-D\tau_\text{sf}(\pm p - 3-\eta_z +8\i\tau_\text{sf}\varepsilon)}/(2D\tau_\text{sf}),\notag\\
\kappa_\pm &= 3 - \eta_z \pm p,\; p = \sqrt{(3-\eta_z)^2 - 64\tau_\text{sf}^2h^2},\notag\\
k &= \sqrt{[2\i\varepsilon - \eta_{xy}/(2\tau_\text{sf})]/D}.
\end{align}
Taking $\eta_z=\eta_{xy}=1$, the above expressions reduce to those of Ref.~\onlinecite{linderUsadel07}. 
Once the $\{f'_{\text{s}},f'_{\text{t}},f'_{\sigma}\}$ have been obtained, one may transform them back to the 
original quantization axis $\parallel \hat{\mathbf{z}}$. If we write
\begin{equation}
	\underline{f} = (f_\text{s} + \mathbf{f}\cdot\underline{\boldsymbol{\tau}})\i\underline{\tau_2}
\end{equation}
one may from Eq. \eqref{eq:anomaleous} identify the vector anomalous Green's function
\begin{equation}
\mathbf{f} = [f_\downarrow-f_\uparrow, -\i(f_\uparrow+f_\downarrow), 2f_\text{t}]/2.
\end{equation}
This is equivalent to the $\mathbf{d}_\vk$-vector formalism described in for instance Ref.~\onlinecite{leggett1975}. 
For future use, we define $f_\pm = f_t\pm f_s$.  Finally, the transformation to the new spin basis is (see 
also Fig. \ref{fig:basis}) 
\begin{equation}
(\mathbf{f}')^\text{T} = \begin{pmatrix}
1 & 0 & 0 \\
0 & \cos\phi & -\sin\phi \\
0 & \sin\phi & \cos\phi \\
\end{pmatrix}
(\mathbf{f})^\text{T}.
\end{equation}

In general, the linearization of the Usadel equation is a valid approximation in the case of a weak proximity effect. This 
may be obtained in two limiting cases: \textit{i)} the barriers have low transparency or \textit{ii)} the transparency is 
perfect (ideal interfaces) and the temperature in the superconducting reservoir is close to $T_c$, such that $\Delta_0$ 
is small. An analytical approach is permissible in both scenarios, with differing boundary conditions. In case \textit{i)}, 
the standard Kupryianov-Lukichev (K-L) boundary conditions \cite{kupluk} are usually employed in the literature, while 
case \textit{ii)} implies continuity of the Green's function and its derivative. In an experimental situation, the barrier 
region can hardly be considered as fully transparent, such that the K-L boundary conditions are more realistic than 
continuity of the Green's function and its derivative. We will therefore employ the K-L bondary conditions in this paper \footnote{It should be noted that the K-L boundary conditions may also be used in the special case of a barrier with perfect transparency, yielding continuity of the Green's function and its derivative.}.
\par
To obtain the anomalous Green's function, we must supplement the general solution in Eq. (\ref{eq:general}) with the 
K-L boundary conditions at three interfaces. At the S/F$_i$ interfaces located at $x=-d_1$ and $x=d_2$, one obtains
\begin{align}
2\gamma d_1 \hat{g}_{F_1} \partial_x \hat{g}_{F_1}|_{x=-d_1} &= [\hat{g}_S(-\chi),\hat{g}_{F_1}]|_{x=-d_1},\notag\\
2\gamma d_2 \hat{g}_{F_2} \partial_x \hat{g}_{F_2}|_{x=d_2} &= [\hat{g}_{F_2},\hat{g}_S(\chi)]|_{x=d_2},\notag\\
\end{align}
We have defined the parameter
\begin{equation}
\gamma = \frac{R_B}{R_F},
\end{equation}
where $R_B$ is the resistance of the barrier region and $R_F$ is the resistance of in the diffusive ferromagnetic regions 
(assumed to be the same for both F$_1$ and F$_2$). For the F$_1$/F$_2$ 
interface, which denotes the separation of the ferromagnetic layers, we assume the resistance to be much smaller 
than at the S/F$_i$ interfaces. Therefore, we model this by continuity of the Green's function and its derivative:
\begin{align}
g_{F_1} = g_{F_2} |_{x=0},\; \partial_x g_{F_1} = \partial_x g_{F_2}|_{x=0}.
\end{align}
Let us comment briefly on the case where $F_1$ and $F_2$ are two domains of a single layer. A domain-wall resistance may 
quite generally be defined as $R_w = R-R_0$, where $R$ and $R_0$ are the electrical resistances with and without a domain
wall (\ie homogeneous magnetization), respectively. When the width of the domain wall increases, $R_w\to0$, and vanishes 
all together when the width of the domain wall is much larger than the Fermi wavelength \cite{kietPRL}. In the present 
paper, we consider an abrupt change in magnetization at the interface of the two domains, corresponding to a very thin 
domain wall, such that one would in general expect a finite contribution to the resistance of the junction. To reduce 
the number of parameters in the problem, however, we \textit{assume} that this resistance is much smaller than at the 
S/F interfaces and effectively set it to zero. In the case where $F_1$ and $F_2$ are separate ferromagnetic layers, 
one may neglect the resistance at the interface between them by assuming a good electrical contact achieved during 
deposition of the layers. 
\par

The $\{f_{2\pm},f_{2\sigma}\}$ anomalous Green's functions are related to a a set of anomalous Green's functions in a 
rotated basis $\{f'_{2\pm},f'_{2\sigma}\}$ via
\begin{align}
f_{2\uparrow} &= \frac{1}{2}[\cos\phi(f'_{2\downarrow} + f'_{2\uparrow}) + 2\i \sin\phi f'_{2,t} + f'_{2\uparrow} - f'_{2\downarrow}],\notag\\
f_{2\downarrow} &= \frac{1}{2}[\cos\phi(f'_{2\downarrow} + f'_{2\uparrow}) + 2\i \sin\phi f'_{2,t} - f'_{2\uparrow} + f'_{2\downarrow}],\notag\\
f_{2,t} &= \cos\phi f'_{2,t} + \frac{\i \sin\phi}{2} (f'_{2\downarrow} + f'_{2\uparrow}).
\end{align}
where $\{f'_{2\pm},f'_{2\sigma}\}$ have the general form as shown in Eq. (\ref{eq:general}). The complete anomalous 
Green's functions in the regions F$_1$ may be written as
\begin{align}\label{eq:f1}
f_{1\pm} &= b_1\e{-q_-x}L_+^\pm(0) + b_2\e{q_-x}L_+^\pm(0) \notag\\
&+ b_3\e{q_+x}L_-^\pm(0) + b_4\e{-q_+x}L_-^\pm(0),\notag\\
f_{1\sigma} &= A_\sigma \e{\i kx} + B_\sigma \e{-\i kx},
\end{align}
while in F$_2$ one finds
\begin{align}\label{eq:f2}
f_{2\pm} &= c_1\e{-q_-x}L_+^\pm(\phi) + c_2\e{q_-x}L_+^\pm(\phi) \notag\\
&+ c_3\e{q_+x}L_-^\pm(\phi) + c_4\e{-q_+x}L_-^\pm(\phi)\notag\\
&+\frac{\i\sin\phi}{2}[(C_\uparrow+C_\downarrow)\e{\i kx} + (D_\uparrow+D_\downarrow)\e{-\i kx}],\notag\\
f_{2\sigma} &= \frac{1}{2}[\e{\i kx}\{\cos\phi(C_\uparrow+C_\downarrow)+ \sigma(C_\uparrow-C_\downarrow)\} \notag\\
&+ \e{-\i kx}\{\cos\phi(D_\uparrow+D_\downarrow) + \sigma(D_\uparrow-D_\downarrow)\} \notag\\
&+ 2\i\sin\phi(c_1\e{-q_-x} + c_2\e{q_-x} +c_3\e{q_+x} + c_4\e{-q_+x})].
\end{align}
Above, we have defined $L_\pm^{\pm'}(\phi) = \cos\phi \pm' \i\kappa_\pm/(8\tau_\text{sf}h)$. Note that Eq. (\ref{eq:f2}) 
reduces to exactly the same form as Eq. (\ref{eq:f1}) for $\phi=0$ (parallel magnetization), as demanded by consistency. 
The remaining task is to determine the 16 unknown coefficients $\{b_i\}, \{c_i\}, \{A_\sigma,B_\sigma,C_\sigma,D_\sigma\}$.
For clarity, we write out the boundary conditions explicitly. At $x=-d_1$, one has $(\sigma=\uparrow,\downarrow)$
\begin{align}\label{eq:bound_d1}
\gamma d_1\partial_x f_{1\pm} = cf_{1\pm} \mp s\e{-\i\chi},\; \gamma d_1\partial_x f_{1\sigma} = cf_{1\sigma},
\end{align}
while at $x=d_2$ we find
\begin{align}\label{eq:bound_d2}
\gamma d_2\partial_x f_{2\pm} &= \pm s\e{\i\chi} - cf_{2\pm} ,\; \gamma d_2\partial_x f_{2\sigma} = -cf_{2\sigma}, \notag\\
\end{align}
Finally, at $x=0$ one obtains
\begin{align}\label{eq:bound_0}
f_{1\pm} &= f_{2\pm},\; \partial_x f_{1\pm} = \partial_x f_{2\pm},\notag\\
f_{1\sigma} &= f_{2\sigma},\; \partial_x f_{1\sigma} = \partial_x f_{2\sigma}.
\end{align}

Inserting Eqs. (\ref{eq:f1}) and (\ref{eq:f2}) into Eqs. (\ref{eq:bound_d1})-(\ref{eq:bound_0}) yields a set of linear equations which may be represented by a $16\times16$ matrix, and the solution for the 16 coeffiecients is found numerically. Once the anomalous Green's functions have been 
obtained, one may calculate physical quantities of interest. In the present paper, we will be concerned with the Josephson current
\begin{align}
	\label{eq:current}
	\mathbf{j}(x) &= -(N_\text{F}eD\hat{\mathbf{x}}/4)\int\text{d}\varepsilon \text{Tr}\{\hat{\rho}_3 	( \hat{g}^\text{R}\partial_x\hat{g}^\text{K} - \hat{g}^\text{K}\partial_x\hat{g}^\text{A} ) \}\notag\\
	&= -(N_\text{F}eD\hat{\mathbf{x}}/2) \int^\infty_{-\infty} \text{d}\varepsilon 	\text{Re}\{M_+(\varepsilon)+M_-(\varepsilon)\notag\\
	&\hspace{0.4in}M_\uparrow(\varepsilon)+M_\downarrow(\varepsilon) \}\times\tanh(\beta\varepsilon/2),
\end{align}
with the definitions ($\sigma=\uparrow,\downarrow$)
\begin{align}\label{eq:jos}
M_{\sigma}(\varepsilon)&=[f_\sigma(-\varepsilon)]^*\partial_x f_{\sigma}(\varepsilon) - f_\sigma(\varepsilon)\partial_x[f_{\sigma}(-\varepsilon)]^*,\notag\\
M_{\pm}(\varepsilon)&=[f_\pm(-\varepsilon)]^*\partial_x f_\mp(\varepsilon) - f_\pm(\varepsilon)\partial_x[f_\mp(-\varepsilon)]^*.
\end{align}
The matrix $\hat{\rho}_3$ in the first line of Eq. \eqref{eq:current} is defined by $\hat{\rho}_3 = \mathrm{diag}(1,1,-1,-1)$. 
The normalized current is defined as 
\begin{align}
	\label{eq:I_c}
	I(\chi)/I_0 = 4|\mathbf{j}(x,\chi)|/(N_\text{F}eD\Delta_0^2),
\end{align}
which is independent of $x$ for $x\in[-d_1,d_2]$ (due to conservation of electrical current). The maximal supercurrent the system can support, known as the \textit{critical current}, is given by $I_\mathrm{c} = I(\frac{\pi}{4})$ in the case of a sinusoidal current-phase relation.

Before proceeding to disseminate our results, we briefly remind the reader (see \eg Refs.~\onlinecite{demler,buzdin}) of the 
qualitative physics that distinguishes S/F proximity structures from S/N systems, and thus gives rise to \eg  0-$\pi$ 
oscillations of the critical current (which will be discussed in detail for our system in Sec. \ref{sec:results}). The 
fundamental difference between the proximity effect in an S/N structure as compared to an S/F structure is that the Cooper 
pair wavefunction acquires a finite center-of-mass momentum in the latter case due to the Zeeman-energy splitting between 
the $\uparrow$- and $\downarrow$-spins constituting the Cooper pair. The finite center-of-mass momentum of the Cooper pair 
implies that the condensate wavefunction in the ferromagnetic region displays oscillations in space, permitting it to change 
sign upon penetrating deeper into the ferromagnetic region. Quite generally, one may write that the Cooper pair wavefunction 
(order parameter) as $\Psi = \Psi_0 \e{-k_sx}$ in an S/N structure while $\Psi = \Psi_0 \cos(k_{f,1}x) \e{-k_{f,2}x}$ in an 
S/F structure, where $\{k_s,k_{f,1},k_{f,2}\}$ are wavevectors related to the decay and oscillation lengths of the 
proximity-induced condensate in the non-superconducting region. 
\par
The fact that the proximity-induced superconducting order parameter oscillates in the ferromagnetic region suggests that the 
energetically most favorable (ground-state) phase difference between the superconducting reservoirs might not always be zero, 
as in an S/N/S junction. For a very thin ferromagnetic layer, $\Psi$ does not change much and there is no reason for why there 
should be an abrupt discontinuity in the phase at one of the F/S interfaces - hence, the system is in the 0-state. If the 
thickness of the ferromagnetic layer is comparable to the oscillation length of $\Psi$ ($\sim 1/k_{f,1}$ in our notation 
above), then $\Psi$ may cross zero in the middle of the ferromagnetic region and display antisymmetric behaviour. This is 
accompanied with a shift of sign of the order parameter in the, say, right bulk superconductor as compared to the left bulk 
superconductor. Under such circumstances, the energetically most favorable configuration corresponds to a phase difference 
of $\pi$ between the superconductors, since $\Delta=\Delta_0$ in one of the superconductors and 
$\Delta=-\Delta_0=\Delta_0\e{\i\pi}$ in the other superconductor. 
\par
This is related to 0-$\pi$ oscillations of the critical current as follows. The energy of the Josephson junction may in the 
tunneling limit be well approximated by
\begin{align}
\varepsilon_J \sim I_{J,0}(1-\cos2\chi),
\end{align}
where $I_0$ contains the magnitude and sign of the critical current while $2\chi$ is the phase difference between the 
superconductors. Now, $I_{J,0}$ is closely related to the proximity-induced condensate wavefunction $\Psi$ in the 
ferromagnetic region, and may change sign depending on for instance the width of the ferromagnetic layer. Depending 
on its sign, the ground-state configuration corresponds to either $2\chi=0$ or $2\chi=\pi$, and the critical current 
supported by the system will change sign depending on which of these phases the system is in (although the critical 
current itself is given by $2\chi=\pi/2$).

\section{Results}\label{sec:results}
Unless otherwise stated, we will fix $h/\Delta_0=10$ and $\gamma=5$ to model a realistic experimental setup with weak ferromagnets (the exchange 
field was estimated to 5-10 meV in Ref.~\onlinecite{sellier}.) The particular choice of $\gamma$ is motivated by the fact that we expect the resistance of the barrier region to exceed the bulk resistance of the ferromagnets, and in addition to a low transparency of the S/F interfaces this clearly suggests that $\gamma \gg 1$. This will ensure a weak proximity effect, as explained in the previous section. In order to reduce the number of free parameters further, we in general set the widths of region F$_1$ and F$_2$ to be equal: $d_1=d_2\equiv d$. The superconducting coherence length $\xi = \sqrt{D/\Delta_0}$ will be used as the unit in which the widths are measured. Where spin-flip scattering is included, we will use the parameter $g = 1/(\tau_\text{sf}\Delta_0)$ as a measure of the strength of this effect.

We have numerically confirmed that the system follows a sinusoidal current-phase relation regardless of the direction of magnetization and all other variable parameters of our system. This allows us to focus on the state that supports the critical current, namely $\chi=\pi/4$. A sinusoidal current-phase relation is moreover what should be expected for such systems with weak proximity effect \cite{current-phase}, and where the bulk superconductors have the same symmetry \cite{Linder_oddfreq}. We have also confirmed that the assumption of weak proximity effects holds by assuring that the value of the anomalous Green's function always obey $|f_{s,t,\sigma}| \ll 1$ for the 
parameter range we consider. To make the computation of the solutions to the Usadel equation numerically stable, we furthermore add a small imaginary term $\i\delta$ to the excitation energy $\varepsilon$, where the value $\delta = 10^{-3}$ has been used. This can be motivated as a way to account for inelastic scattering processes \cite{dynes}, interpreting the term as the inverse (positive) 
quasiparticle lifetime.

\subsection{Zero temperature}
\begin{figure}[h!]
\centering
\resizebox{0.48\textwidth}{!}{
\includegraphics{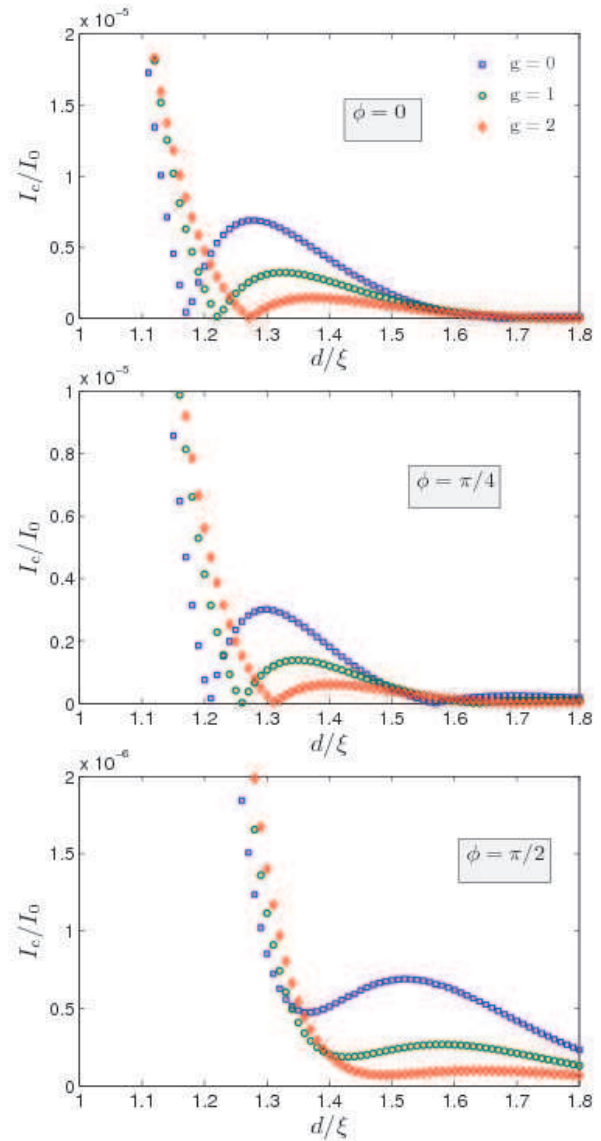}}
\caption{(Color online) Plot of the width-dependence of the critical current for several values of the spin-flip scattering 
rate, which is here taken to be uniaxial in spin space. We have defined the dimensionless parameter $g=1/(\tau_\text{sf}\Delta_0)$ 
as a measure of the spin-flip scattering rate. At $d/\xi = 1.0$, the system is in the 0 state.}
\label{fig:I_vs_d_uni}
\end{figure}

First, we will consider the case where the temperature is fixed to zero unless otherwise stated, \ie the calculations are made 
with $T/T_c = 0.001$. The critical current as a function of the junction width $d$ is shown in Fig. \ref{fig:I_vs_d_uni} for isotropic and uniaxial spin-flip scattering, respectively. Considering first the parallel 
case $\phi = 0$, the well known 0-$\pi$-oscillations are reproduced, where the current change sign for certain values of 
$d$. It should however be noted that we have chosen to always plot the critical current as a positive quantity, as defined 
by Eq. \eqref{eq:I_c}, because this is what is most commonly measured in experiments \cite{ryazanovPRL01,buzdin}. We have 
confirmed that the oscillations are almost exactly periodic for the parameter range considered here. Increasing the effect 
of spin-flip scattering tends to move the transition points between the 0 state and the $\pi$ state towards higher values 
of $d$. Throughout our investigations, we find no significant difference between isotropic and uniaxial spin-flip scattering for the width dependence 
of the critical current. Thus, we consider only uniaxial spin-flip scattering whenever the role of magnetic impurities is studied.

\begin{figure}[h!]
\centering
\resizebox{0.48\textwidth}{!}{
\includegraphics{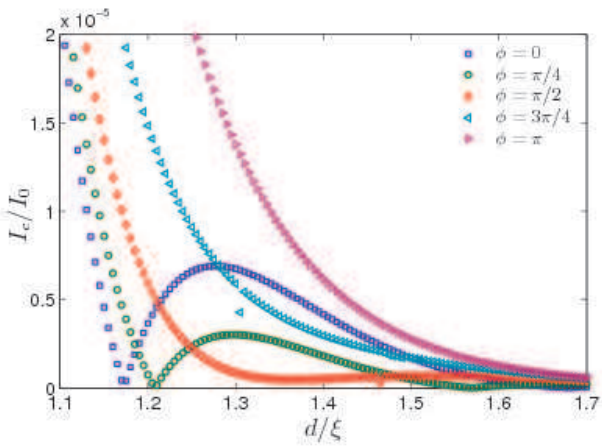}}
\caption{(Color online) Plot of the width-dependence of the critical current for several values of the misorientation angle $\phi$, with spin-flip scattering turned off for simplicity ($g=0$).}
\label{fig:I_vs_d_5xphi}
\end{figure}

The subplots of Fig. \ref{fig:I_vs_d_uni} show how increasing the relative angle of magnetization 
to $\phi = \pi/4$ shifts the first transition points to the right and the second to the left, reducing the width region in 
which the system is in the $\pi$ state. For $\phi = \pi/2$, the oscillations have ceased entirely, leaving the junction in 
the 0 state for all values of $d$. These effects are shown more clearly in Fig. \ref{fig:I_vs_d_5xphi}, which also shows 
that the oscillations do not return for $\phi \in [\pi/2,\pi]$. This can be expressed as a critical misorientation angle 
$\phi_c \lesssim \pi/2$ over which the $\pi$ state is not realizable, which is in agreement with the findings of 
Ref. ~\onlinecite{crouzy1}. Ref.~\onlinecite{crouzy1} claims that $\phi_c \equiv \pi/2$ independent of parameters as 
long as the system is near the critical temperature, $T/T_c\simeq 1$. We find $\phi_c$ to have a somewhat lower value 
$\phi_c \simeq 0.46\pi < \pi/2$ in the present case of $T/T_c\ll 1$, but we will show in Sec. \ref{sec:finiteT} how 
this value approaches $\pi/2$ for increasing temperatures, and how it changes for other values of the exchange field 
than our particular choice of $h/\Delta_0=10$. In light of the discussion concerning the qualitative physics involved 
in a S/F/S proximity structure, it seems reasonable to suggest that the vanishing of the 0-$\pi$ transitions are 
directly linked to a strong modification of the Cooper-pair wavefunction oscillation length inside the ferromagnet, 
which renders the $\pi$-state inaccessible. Possible explanations will be discussed further in Sec. \ref{sec:disc}. 

\begin{figure}[h!]
\centering
\resizebox{0.48\textwidth}{!}{
\includegraphics{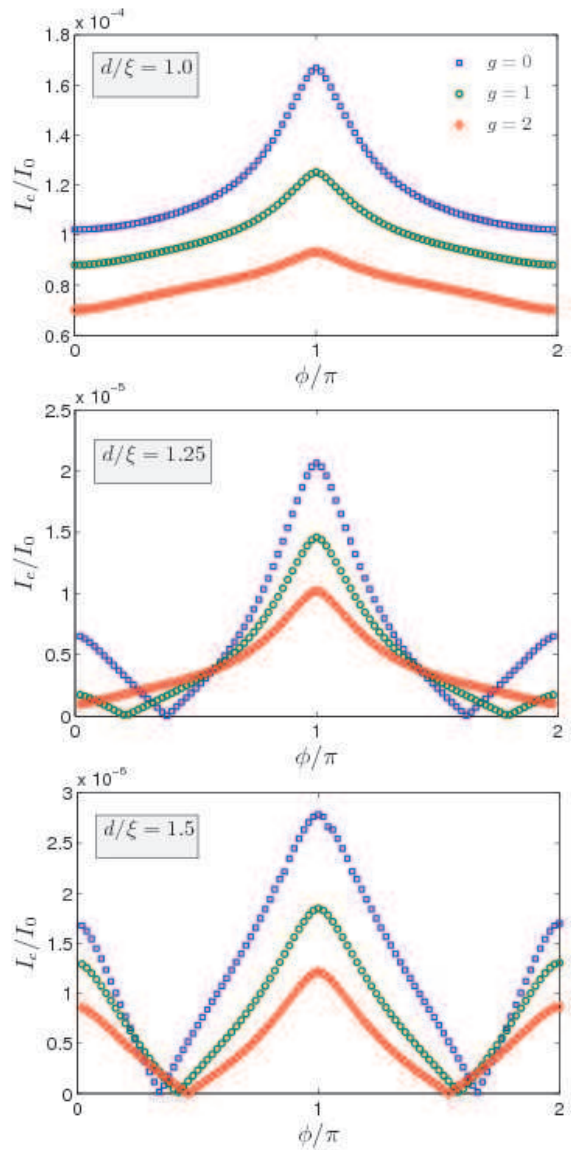}}
\caption{(Color online) Plot of how the critical current is affected by a change in the relative orientation $\phi$ of the 
magnetizations for several values of the spin-flip scattering rate, which is here taken to be uniaxial in spin space.}
\label{fig:I_vs_phi_uni}
\end{figure}

As stated in the introduction, several works have contrasted the cases of parallel and antiparallel orientations, while the 
intermediate angles have not been studied thoroughly (see, however, Refs.~\onlinecite{barash,crouzy1}). We seek to remedy 
this by first presenting in Fig. \ref{fig:I_vs_phi_uni} the dependence of the critical current 
on the misorientation angle. The three junction widths $d/\xi = \{1.0, 1.25, 1.5\}$ are chosen somewhat arbitrarily from 
the available range of values, but illustrate adequately the conditions for appearance of 0-$\pi$-oscillations. First, we 
confirm the obvious fact that the critical current should be symmetric with respect to $\phi = \pi$. Therefore we will from 
now on only consider the interval $\phi \in [0, \pi]$, the maximum of the misorientation angle being the antiparallel 
orientation. Next, we see that no 0-$\pi$ oscillations appear upon varying the misorientation angle for $d/\xi = 1.0$ 
fixed, which agrees with the observation from the previous figures that the junction appears to be in the 0 state for 
all angles at this junction width. For $d/\xi = \{1.25, 1.5\}$ however, the junction starts out in the $\pi$ state for 
the parallel orientation, and we can see that a transition takes place to the 0 state for some angle $\phi < \pi/2$. This 
is in agreement with the result of Ref.~\onlinecite{barash} that a nonmonotonic dependence of the critical current on 
$\phi$ occurs when the $\pi$ state is the equilibrium state of the junction for $\phi = 0$, and a similar statement 
was also made in Ref.~\onlinecite{pajovic}.

\par

A small effect of spin-flip scattering which may be 
mentioned here, is that for $\phi > 0$ it may give the appearance of an effectively lowered misorientation angle with 
regard to the shift in the $d$ values for 0-$\pi$ crossover. As a result, for angles just above the critical angle, 
an increase of $g$ may trigger the transition from complete absence of the $\pi$ state to 0-$\pi$ oscillations.

\begin{figure}[h!]
\centering
\resizebox{0.45\textwidth}{!}{
\includegraphics{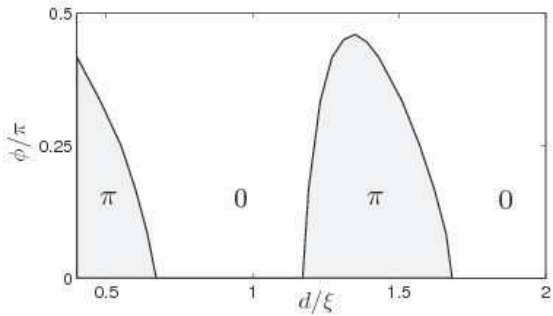}}
\caption{Phase diagram of the Josephson junction for zero temperature, showing the regions occupied by the 0 state and the 
$\pi$ state in width-misorientation parameter space. For the region given by $\phi \in [\pi/2,3\pi/2]$, the $\pi$ state is 
completely absent. We have  set the spin-flip scattering strength to zero, $g=0$.}
\label{fig:phase_diag_d-phi}
\end{figure}

The evolution of the critical current for variable $d$ and $\phi$ described in the previous paragraphs can be condensed to 
the phase diagram shown in Fig. \ref{fig:phase_diag_d-phi}. If only the sign of the Josephson current is of interest, each 
of the plots of $I_\mathrm{c}(d)$ may be thought of as a horizontal sweep through the phase diagram for some fixed $\phi$, 
while every plot of $I_\mathrm{c}(\phi)$ is represented by a vertical sweep for some fixed $d$. As seen, the $\pi$-state becomes impossible above a critical angle $\phi_c$ for the present case of $T\to0$. We will contrast this with the finite-temperature case in the next section.

%First, increased spin-flip scattering not only shifts the diagram to the right, but it is also found to widen the region of 
%each phase slightly, thus making the shift between the phase diagram with $g > 0$ and that with $g = 0$ progressively larger 
%for increasing $d$. This is explained by the fact that spin-flip scattering increases the oscillation length of the Green's
%functions \cite{linderUsadel07}, resulting here in an increased period of the 0-$\pi$ oscillations in the current. Second, 
%closer examination shows that spin-flip-enhancement of the current, $I_c(g > 0) > I_c(g = 0)$, is the case in regions close 
%to the boundary between the 0 phase and the $\pi$ phase. Upon increasing $d$ in Fig. \ref{fig:phase_diag_d-phi}, one enters 
%this region some time before hitting the boundary, but leaves it almost immediately after entering the next phase. In fact, 
%considering also the sign of the critical current, we find that $I_{J,0}(g > 0) > I_{J,0}(g = 0)$ holds in regions almost 
%coinciding with those of the $\pi$ state, but shifted to the left. The fundamental reason for this is that the spin-flip 
%rate affects the oscillation length of the Cooper pair wavefunction in the ferromagnetic region, which allows the %function to become enhanced compared to the $g=0$ case although its magnitude is in general suppressed \cite{linderUsadel07}. 
%The fact that the critical current is calculated as the absolute value $I_c = |I_{J,0}|$ furthermore explains why the region of 
%enhancement ends immediately after the transition from one phase to another. 

\subsection{Finite temperature} \label{sec:finiteT}

\begin{figure}[h!]
\centering
\resizebox{0.45\textwidth}{!}{
\includegraphics{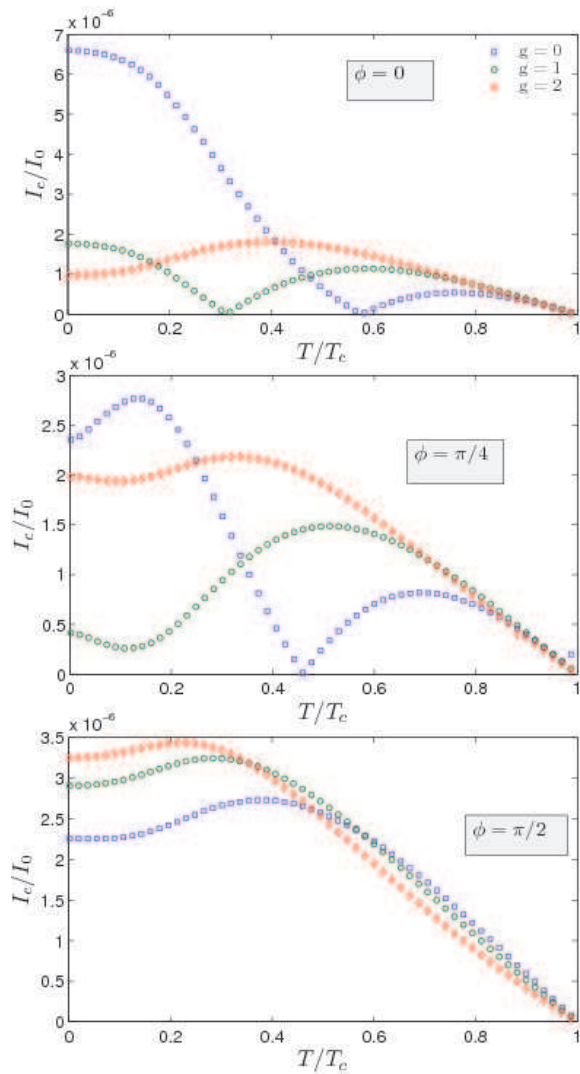}}
\caption{(Color online) Plot of the temperature dependence of the critical current for several values of the spin-flip 
scattering rate, which is here taken to be uniaxial in spin space. The junction width is given by the value $d/\xi = 1.25$. }
\label{fig:I_vs_T_uni}
\end{figure}

We proceed by considering the dependence of the Josephson current on the temperature. The superconducting electrodes were assumed to be conventional superconductors unaffected by the ferromagnetic layers, so the standard BCS temperature dependence of the superconducting gap will be employed:
\begin{equation}
	\Delta(T) = \Delta_0 \tanh \left( 1.74 \sqrt{T_c/T - 1} \right).
\end{equation}

\begin{figure}[h!]
\centering
\resizebox{0.48\textwidth}{!}{
\includegraphics{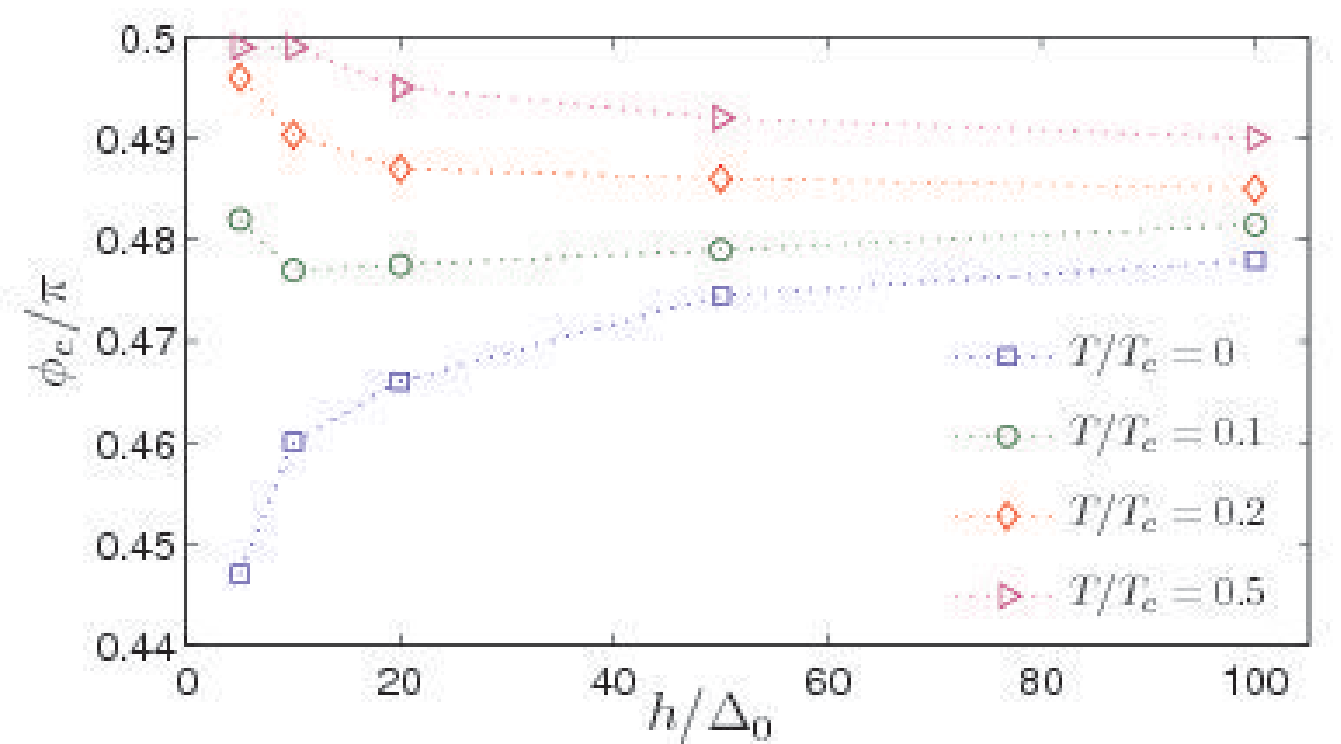}}
\caption{(Color online) Plot of the critical angle $\phi_c$ at which the 0-$\pi$ oscillations disappear, as a function of 
the exhange splitting $h$, for a number of temperatures. Spin-flip scattering is neglected for simplicity, $g=0$.}
\label{fig:phi_vs_h}
\end{figure}

To illustrate how the critical angle $\phi_c$ for 0-$\pi$-oscillations depends upon increasing the temperature, including also 
the dependence on the exchange field, we plot in Fig. \ref{fig:phi_vs_h} the critical angle as a function of the exchange 
splitting $h$ for several values of $T$. As seen, $\phi_c$ remains less than $\pi/2$ up to $h/\Delta_0 \simeq 100$ in the 
$T\to 0$ limit. However, increasing the temperature only slightly to $T/T_c=0.2$, we see that $\phi_c$ rapidly approaches 
$\pi/2$. The trend is the same upon increasing the temperature even further, indicating the limit of $\phi_c = \pi/2$ for 
arbitrarily high values of $h$ as $T \simeq T_c$. We conclude therefore that the critical angle above which 0-$\pi$ oscillations 
cease to exist is equal to $\pi/2$ as long as the temperature is high ($T/T_c \simeq 1$). However, for low temperatures and 
weak exchange fields, we find that $\phi_c$ deviates noticeably from $\pi/2$. 

Also by varying the temperature parameter may 0-$\pi$ oscillations be found in the system, as shown in Figs. \ref{fig:I_vs_T_uni}. This follows as a natural result if the critical values of junction width at which 0-$\pi$ transitions 
were found in the preceding section are temperature-dependent. One difference from the plots of $I_\mathrm{c}$ as a function of 
$d$ is the existence of no more than one transition point (for each value of $g$), before the Josephson current necessarily 
vanishes at $T = T_c$. For increasing $\phi$, this transition point moves leftwards until vanishing at $T = 0$ at some critical 
angle. By considering the dependence on misorientation angle more carefully, we found that this critical angle, over which the 
0-$\pi$ oscillations disappear for $I_\mathrm{c}$ as a function of $T$, differs from the corresponding critical angle for 
$I_\mathrm{c}$ as a function of $d$. This may be explained by going back to the phase diagram in Fig. \ref{fig:phase_diag_d-phi} 
and noticing that (for $T/T_c \approx 0$) the maximum $\phi$ for the $\pi$ phase region corresponds to a junction width 
$d/\xi \approx 1.35$ (up to periodicity). For any other junction width, \eg $d/\xi = 1.25$ as used in the figures, the 
$\pi$ state will be unrealizable at $T = 0$ for even lower values of $\phi$. This results in the inequality that $\phi_c$ 
for thermally induced oscillations is always less than or equal to $\phi_c$ for width-induced oscillations.

Another point on which the thermally induced 0-$\pi$ oscillations differs from those obtained by varying the junction width 
or the misorientation angle, is the remarkably stronger dependence on the spin-flip scattering rate. Increasing $g$ shifts 
the transition point significantly to the left, and furthermore strongly influence the ratios $I_\mathrm{c}(g > 0)/I_\mathrm{c}(0)$. 
Similar findings were reported in Ref.~\onlinecite{lts}.

%In a similar manner as in the case of parameters $d$ and $\phi$, it is also instructive to construct a $d$-$T$ phase diagram 
%for the system. This is done in Fig. \ref{fig:phase_diag_d-T} for the parallel orientation. Observing this phase diagram, we 
%note that only for certain small regions of junction width are thermal 0-$\pi$ oscillations attainable, since the transition 
%points for width-induced oscillations varies so weakly with $T$. Building on this argument, one can also see that our result 
%agrees well with the statement of Radovic \etal that the thermally induced 0-$\pi$ transitions are only possible when the 
%system is close to a transition point at $T = 0$ \cite{radovic}.

\begin{figure}[h!]
\centering
\resizebox{0.45\textwidth}{!}{
\includegraphics{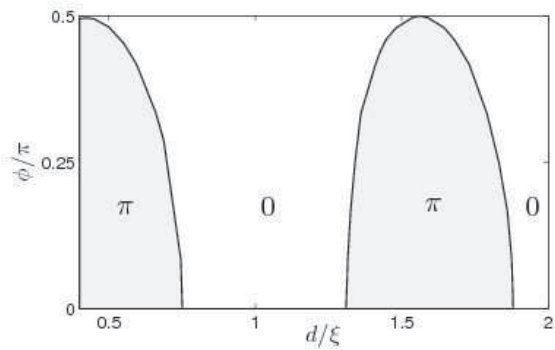}}
\caption{Phase diagram of the Josephson junction for temperature $T/T_c = 0.5$, showing the occupation of the 0 state and 
the $\pi$ state in width-misorientation parameter space in a similar manner as Fig. \ref{fig:phase_diag_d-phi} does for 
zero temperature. For the region given by $\phi \in [\pi/2, 3\pi/2]$, the $\pi$ state is completely absent. We have here 
set the spin-flip scattering strength to zero, $g=0$.}
\label{fig:phase_diag_d-phi_T05}
\end{figure}

So far, we have not considered the dependence of the critical current on the misorientation angle while simultaneously going 
away from the limiting case of $T = 0$. In principle the phase diagrams presented might readily be generalized to a 
3-dimensional $d$-$\phi$-$T$ phase diagram, but we justify the omittance of this by arguing that the phase diagram of the 
system does not contain many interesting new features not already contained in the 2-dimensional projection presented here. However, 
as is clearly visible in the phase diagram for $T/T_c = 0.5$ as shown in Fig. \ref{fig:phase_diag_d-phi_T05}, the 
critical angle is exactly $\phi_c = \pi/2$, in full agreement with the analysis done in Ref. ~\onlinecite{crouzy1}. Another 
development as $T$ increases is the $\pi$-region obtaining a more symmetric shape, also this in agreement with 
Ref. ~\onlinecite{crouzy1}.

\subsection{Enhancement effect}

\begin{figure}[h!]
\centering
\resizebox{0.46\textwidth}{!}{
\includegraphics{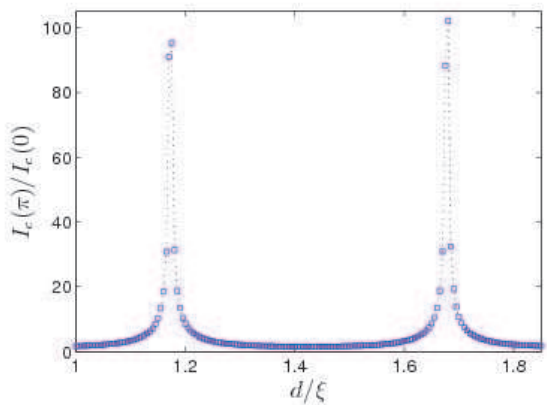}}
\caption{(Color online) Plot of the ratio between the critical current in the antiparallel $[I_\mathrm{c}(\phi = \pi)]$ 
and parallel $[I_\mathrm{c}(\phi = 0)]$ orientation as a function of width $d$. We have set $g=0$, but the behaviour is 
qualitatively identical for $g>0$. }
\label{fig:enhancement}
\end{figure}

As was seen in Fig. \ref{fig:I_vs_phi_uni}, there is a significant difference between the 
current in the parallel configuration $\phi = 0$ and the antiparallel $\phi = \pi$. The ratio between these two critical 
currents, $I_\mathrm{c}(\phi = \pi) / I_\mathrm{c}(\phi = 0)$, is plotted as a function of the junction width in 
Fig. \ref{fig:enhancement}. The fact that one always observes 0-$\pi$ oscillations for varying $d$ in the parallel 
case, but never in the antiparallel, leads to a divergence of the ratio at certain values of the junction width, 
since $I_\mathrm{c}(0)$ drops to zero at this transition point while $I_\mathrm{c}(\pi)$ remains finite for all 
values of $d$. We note also that one will always have $I_\mathrm{c}(\pi) > I_\mathrm{c}(0)$, but the critical current 
is a monotonously increasing function of $\phi$ up to $\phi = \pi$ only in the case that the system is in the 0 state 
for $\phi = 0$. 

This enhancement of the Josephson current by switching the direction of magnetization may possibly be utilized in a 
device for controlling the magnitude of the current, if the junction is tuned to the vicinity of a transition point. 
A similar effect was mentioned by Golubov \etal \cite{golubovJETP}, who considered the exchange field $h$ as the 
variable parameter, but to our knowledge it has not yet been pointed out how this effect may be applied by tuning 
the junction width or the temperature.

\section{Discussion}\label{sec:disc}

Above, we have neglected the spatial depletion of the superconducting order parameter near the S/F interfaces. This 
approximation is expected to be excellent in the case of a low-transparency interface. \cite{bruder} Moreover, it is 
well-known that a magnetic flux threading in a Josephson junction in general gives rise to a Fraunhofer modulation of 
the current as a function of the flux. \cite{sudbo_book} We here neglect this modification by assuming that the flux 
constituted by the ferromagnetic region is sufficiently weak compared to the elementary flux quantum. This is the 
case for either a small enough surface area or weak enough magnetization, but neither of these preclude the possibility 
of having an appreciable energy \textit{exchange splitting} between the majority and minority spin bands.
\par
In the limit of antiparallel orientation, the $\pi$ state will become disallowed because the effect of the ferromagnetic layers cancels, effectively giving a S/N/S junction. However, remembering the symmetry requirements around $\phi = \pi$ and the possibility that also partial cancellation is sufficient to render the sign change in $I_{J,0}$ impossible, we realize that 0-$\pi$-oscillations may vanish for two intermediate angles $\phi = \pm \phi_c$ with $0 < \phi_c < \pi$. The partial cancellation of the exchange fields commences at $\phi \geq \pi/4$, which may provide a clue as to why the critical angle is always in the vicinity of $\phi=\pi/4$ as seen in Fig. \ref{fig:phi_vs_h}. Note that although the 0-$\pi$ oscillations vanish above the critical angle $\phi_c$, it is evident from \eg Fig. \ref{fig:I_vs_d_uni} that  the critical current does not decay monotonously even for $\phi > \phi_c$ as it would have in an S/N/S junction.  
\par
One interesting observation in our study is that although the critical angle varies, we never find $\phi_c > \pi/2$. This means that any choice of parameters that brings us away from the limit considered in Ref. ~\onlinecite{crouzy1} seems to lower $\phi_c$ but never increase it. 
A conjecture which may shed some more light on this phenomenon, is that the magnetization component of the F$_2$ layer perpendicular to the magnetization of the F$_1$ layer can be viewed as an additional effective spin-flip scattering effect. It is known\cite{linderUsadel07} that sufficiently strong spin-flip scattering may remove the oscillations in the anomalous Green's function entirely, thereby inhibiting 0-$\pi$ oscillations. This effect will be at its strongest for $\phi \rightarrow \pi/2$, which may account for the somewhat surprising fact that $\phi_c$ always remains close to $\phi = \pi/2$. Combined with the effective cancellation of the magnetization described in the previous paragraph, this also serves as a possible explanation why the $\pi$-state is forbidden for the orientations $\pi/2 <\phi < \pi$. To gain further understanding of the phase diagram of multilayer SFS junctions, we suggest to extend the study to a trilayer model similar as that studied in Ref. ~\onlinecite{trilayer}, but where the three layers have equal thickness. If one fixes the middle layer and varies the orientation of the leftmost and rightmost layers, the possible existence and value of a critical misorientation angle would give some hints to the origin of this phenomenon also in our bilayer system.
\par
\begin{figure}[h!]
\centering
\resizebox{0.48\textwidth}{!}{
\includegraphics{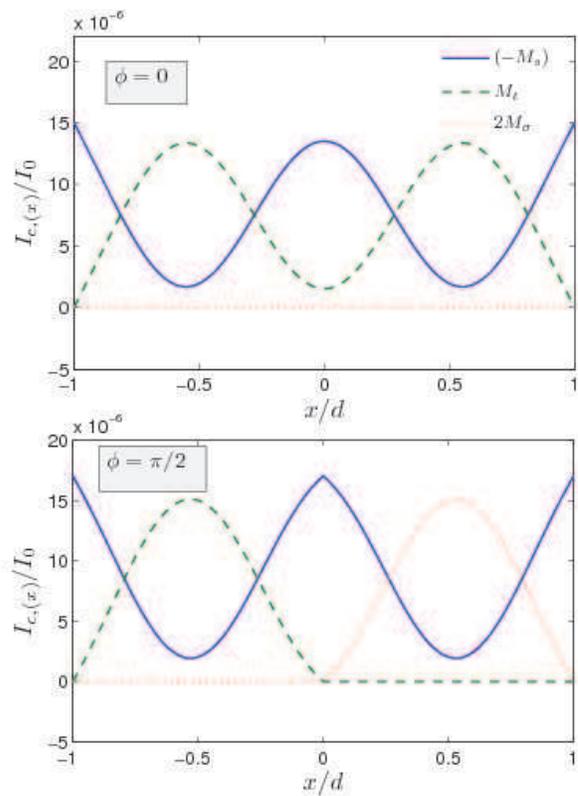}}
\caption{(Color online) Spatial dependence through the ferromagnet for the separate contributions to the critical current, 
as decomposed in Eq. \eqref{eq:decompose}. $I_{\mathrm{c},(x)}$ refer to $I_{\mathrm{c},s}$, $I_{\mathrm{c},t}$ and 
$I_{\mathrm{c},\sigma}$, with contributions to the integrand from the components $[-M_{s}(\varepsilon)]$, $M_{t}(\varepsilon)$ 
and $2M_{\sigma}(\varepsilon)$, respectively. The current was evaluated for a junction width given by $d/\xi = 1.0$ and 
temperature given by $T/T_c = 0.4$, and spin-flip scattering is neglected for simplicity ($g=0$).}
\label{fig:I_vs_x}
\end{figure}

Finally, we would like to present a decomposition of the current to serve as the basis for a discussion on the long 
range contributions to the Josephson current. It can be shown easily from the formula for the Josephson current in 
Eq. \eqref{eq:current} that one may rewrite $M_+ + M_- = M_t - M_s$ where $M_s$ and $M_t$ is expressed exclusively 
by the components $f_s$ and $f_t$ of the Green's function, respectively. It can furthermore be shown that for the 
ESP components one gets $M_\uparrow = M_\downarrow \equiv M_\sigma$. One may in this manner decompose the current as
\begin{align}
	\label{eq:decompose}
	I_\mathrm{c}/I_0 &= (I_{\mathrm{c},s} + I_{\mathrm{c},t} + I_{\mathrm{c},\sigma})/I_0 = \int^\infty_{-\infty} \text{d}\varepsilon \text{Re}\{ [-M_s(\varepsilon)]\notag\\
 	&+ [M_t(\varepsilon)] + [2 M_\sigma(\varepsilon)] \}\times\tanh(\beta\varepsilon/2).
\end{align}
While the total current is easily shown in our framework to be constant throughout the junction, the separate components 
given above need not be, and the spatial dependence of each contribution is plotted in Fig. \ref{fig:I_vs_x} for selected 
parameter values. For parallel orientation, one naturally finds that the ESP correlations do not contribute to the current 
at all. For $\phi = \pi/2$, where the ESP contribution is naively expected to be at its maximum, we find that the OSP 
triplet contribution however, is exactly zero throughout $\mathrm{F}_2$. That $M_t$ and $2M_\sigma$ seem to change roles 
at $x = 0$ can be explained in a natural way by remembering that ultimately, the quantization axis was taken to be 
$\parallel \hat{\mathbf{z}}$ for all $x$. For a quantization axis $\parallel \hat{\mathbf{y}}$, however, the components 
considered as ESP in the former case would here correspond to OSP components, having spin projection $S_y = 0$.
\par

\par
The above argument may be used to to clarify a point regarding the contribution to the current from LRTC. Ref.~\onlinecite{trilayer} 
claims that a long range component of the critical current does not appear in a SF$_1$F$_2$S structure even with noncollinear 
magnetizations, which seems at odds with our observation in Fig. \ref{fig:I_vs_x} of a non-zero ESP component to the current 
in $\mathrm{F}_2$. It is however important to maintain the distinction between the ESP contribution to the current and a LRTC 
contribution. As explained in the case of $\phi = \pi/2$, the total triplet contribution is equal for the two magnetic layers, 
but appears as an ESP component in $F_2$ only because of the choice of quantization axis. Therefore, the ESP contribution in 
$\mathrm{F}_2$ is equivalent to the OSP contribution in $\mathrm{F}_1$, and thus cannot be regarded as a true long range 
component. A long range Josephson effect is defined by the absence of the exchange field in the exponent for the relevant 
Green's functions, making its decay length in a ferromagnet comparable to that of a normal metal. Inspecting 
Eqs. \eqref{eq:f2} and \eqref{eq:q_k}, we see that this certainly is not the case for $\phi = \pi/2$.
\par
The discussion concerning the different contributions to the Josephson current may also hold an important clue concerning 
the 0-$\pi$ oscillations. When inspecting the symmetry components separately, we observed that there can be 0-$\pi$-oscillations of both the singlet and the triplet components simultaneously. If this is generally the case, one idea is to investigate for which parameters these two contributions act constructively and for which they act destructively. In this way the relative interplay of the different symmetry components of $I_c$ may offer an explanation of the behaviour of the critical angle $\phi_c$ for which the 0-$\pi$ oscillations of the total critical current disappear.

\section{Summary}\label{sec:summary}

In conclusion, we have investigated the 0-$\pi$ oscillations of the critical current in a diffusive SF$_1$F$_2$S Josephson 
junction with noncollinear magnetization, where the effects of noncollinearity and spin-flip scattering have been studied 
in particular. The introduction of spin-flip scattering does not change the Josephson current dramatically, so the phase diagrams presented above for zero spin-flip scattering would therefore not be qualitatively 
changed much by setting $g > 0$. Also, comparing isotropic and uniaxial spin-flip scattering, we found that the effect 
of these was very similar both qualitatively and quantitatively.
\par
Oscillations of the critical current for varying junction width disappear when the relative angle of magnetization passes a 
critical value $\phi_c$, making the $\pi$ state unattainable regardless of choice of the other parameters. This critical 
angle equals $\pi/2$ in the limit of relatively high temperatures $(T/T_c\simeq 1)$, but is lowered below $\pi/2$ when 
the temperature is low \textit{and} the exchange field is small simultaneously. These dependencies on the various parameters can rather easily be read out of phase diagrams of the kind we have presented. A straight-forward analytical approach to the behaviour of the critical angle is challenging due to the many variables in our system, but we have discussed several ways by which its origin can be better understood.
\par
With regard to the effect of finite interface transparencies, we have not found any signs throughout our investigations implying that the results of Ref.~\onlinecite{crouzy1} change in any significant way. Our mapping of the relevant parameter regimes 
does however serve as a starting point for looking for new interesting effects that may appear upon varying the 
parameters kept fixed in our case, \ie the transparency $\gamma$ and the exchange field $h$ in particular. A natural 
course for a continuation of this work would be expanding the system from a bilayer ferromagnet to a trilayer, 
similar to the Josephson junction considered in Ref. \onlinecite{trilayer}, where the relevant parameter regime for 
a significant LRTC contribution to the Josephson current was found. It might be interesting to see how LRTC manifests in our framework of symmetry components to $I_c$, and to investigate for what region in parameter space the $\pi$ state can be realized in such a system.

\acknowledgments

J.L. thanks T. Yokoyama and M. Blamire for useful discussions. J.L. and A.S. were supported by the Research Council of Norway, 
Grants No. 158518/432 and No. 158547/431 (NANOMAT), and Grant No. 167498/V30 (STORFORSK).

\end{document}